\documentclass[reprint,superscriptaddress,amsmath,amssymb,aps]{revtex4-2}

\usepackage{bm}
\usepackage{float}
\usepackage{graphicx}
\usepackage[usenames,dvipsnames]{color}
\usepackage[normalem]{ulem}
\usepackage[svgnames]{xcolor}
\usepackage{bm}
\usepackage{multirow}
\usepackage{titlesec}
\usepackage[utf8]{inputenc}
\usepackage{ragged2e} 
\begin{document}

\title{The often overlooked role of sound in liquid thermal transport}
\author{Kamran Behnia}
\email{kamran.behnia@espci.fr}
\affiliation{Laboratoire de Physique et d'\'Etude des Mat\'eriaux \\ 
(ESPCI - CNRS - Sorbonne Universit\'e), PSL Research University, 75005 Paris, France}
\date{\today}

\begin{abstract}
Formulated a hundred years ago, the Bridgman formula for thermal conductivity of liquids postulates a proportionality between sound velocity and thermal conductivity. This designates phonons as heat carriers in a liquid.  Its empirical success has been under-appreciated and its conceptual implications overlooked. Here, we review its origin,  its links to hydrodynamic Euler equations and to the fundamental paradox surrounding the parabolic heat equation and  its fate across the quantum-classical crossover. 
\end{abstract}
\maketitle
\section{Introduction}
The Bridgman formula of liquid thermal conductivity \cite{Brigman1923} was originally published in 1923:

\begin{equation}
\kappa = rk_Bv_sn^{2/3}.
\label{Bridg}
\end{equation}

Here, $r$ is a dimensionless numerical factor of the order of unity, $v_s$ is the sound velocity and $n$ is the particle density of the liquid. While the formula  has stimulated research for more than a century \cite{Rao,Viswanath,Ohtori2014,Khrapak2021,Qing2020,Zhao2021,KHRAPAK2023121786,Khrapak2024}, it is not widely known by quantum transport experts and is not mentioned in prominent condensed matter textbooks \cite{ziman1960electrons,ziman1972,ashcroft_solid_1976,chaikin2000principles,abrikosov1988fundamentals}. 

The present author, after almost four decades of studying thermal transport in solids, first learned about the Bridgman formula when casually browsing  a book dedicated to transport phenomena \cite{bird2006transport}.

This paper  is dedicated to Kostya Trachenko's memory, a foremost pioneer of the phonon theory of liquids \cite{Bolmatov2012,Trachenko_2016,trachenko2023theory}. It presents a short discussion of the Bridgman formula, its origin, its experimental relevance, and its implications. By no means, it can be considered an exhaustive review of thermal transport in liquids. The aim is to explore the surprising efficiency of this formula, in contrast to what the authors of  \textit{Statistical mechanics of nonequilibrium liquids} write:  ``Thermal conductivity has proven to be one of the most difficult transport coefficients to calculate'' \cite{evans2008statistical}.
In addition to recalling what has been  previously published, the paper contains new elements including a discussion of thermal diffusivity in the three states of matter and of the hydrodynamic origin of the Bridgman formula, as well as its link to wave-like propagation of heat beyond the diffusive picture.

We will begin by recalling the significance of particle correlations in liquids, which support the fundamental assumption that collective excitations \cite{Hubbard_1969} can exist in a fluid. We will see that direct experimental evidence for phonons is sparse yet convincing.  

After presenting the  data for the three phases of water, which show the experimental relevance of the formula, we will see the hydrodynamic origin of the formula, starting from Euler's equations. The presence of sound velocity, which  implies a crucial role played by sound in the travel of heat, is a feature deserving attention, given its implications beyond the case of classical liquids. 

Finally, we will discuss the implications for quantum liquids. The quantum-classical crossover is not expected to obliterate these collective excitations. Therefore, it is reasonable to expect heat (and perhaps charge) is not exclusively carried by quasi-particles.  This approach has already been demonstrated to be fruitful in the case of $^3$He \cite{Behnia2024}.
\begin{figure}
 \centering
\includegraphics[width=0.45\textwidth]{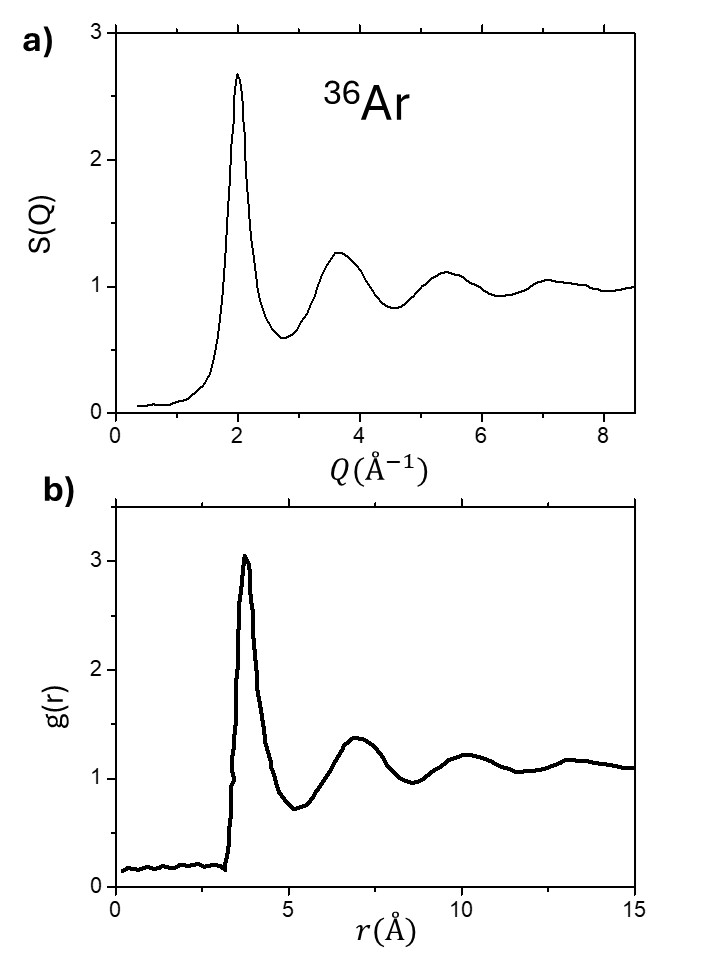}
 \caption{a) Structure factor, $S(Q)$, in liquid argon. b) Radial distribution function, $g(r)$, the Fourier transform of $S(Q)$. The oscillations imply that the presence or absence of an atom at $r=0$ alters the probability of the presence of another atom at $r\neq 0$. This leads to collective phenomena absent in gases.}
\label{fig1}
\end{figure}

\section{Particle correlations and phonons in liquids}
In their remarkable and influential textbook, Chaikin and Lubensky \cite{chaikin2000principles} start by reminding their readers that a gas or a simple liquid is characterized by full translational and rotational symmetry. The average density is uniform with no preferred direction or periodicity. Therefore, density-density correlations decay to a constant at long distances. A crystalline solid, by contrast, spontaneously breaks both continuous translational and rotational symmetry. The density becomes a periodic function of position, defined by a set of reciprocal lattice vectors, and this periodicity persists even at arbitrarily long distances. The main point is that instead of distinguishing the three phases by how tightly packed they are (which makes liquids  closer to solids), it is more profound to examine the hierarchy of broken symmetries (which makes liquids akin to gases).

The approach  sets the stage to introduce liquid crystals \cite{de1993physics} as intermediate cases, in which some symmetries break while  others are preserved, and this can be elegantly contrasted with simple isotropic liquids. However, in the same introductory chapters \cite{chaikin2000principles}, the authors inform the readers that particle-particle correlations, absent in a perfect gas (and negligible in any real gas), are significant in liquids. This makes them akin to solids rather than to gases. Indeed, in a gas the presence of a particle at one place has almost no bearing on the presence or absence of  other particles elsewhere. This is not the case of condensed matter (liquids and solids). From the presence of one particle somewhere, one can deduce information about the presence of other particles. The only difference is that in liquids this information decays with distance but in crystalline solids it does not.

The concept of pair correlation function, $g(r)$ (also called radial distribution function when the fluid is isotropic), captures this  point. $g(r)$ represents the probability density of finding a particle at distance $r$ from a reference particle, relative to that of an ideal (uncorrelated) gas at the same density. 

Yarnell and collaborators \cite{Yarnell1973} used neutron scattering to determine the structure factor $S(Q)$ of liquid argon at T= 85 K. The result is shown in Figure \ref{fig1}a. From the structure factor, they  derived $g(r)$ after performing a Fourier transform (Figure \ref{fig1}b). Oscillations periodic with a decaying amplitude are visible, the period of extracted oscillations in real space corresponds to the average distance between Ar atoms. Thus, as one moves farther away from the central atom, the probability of finding another atom varies periodically and this variation remains detectable up to the fourth period. Interestingly, such a structure for $g(r)$ can be reproduced by replacing the Lennard-Jones interaction with a hard-sphere interaction \cite{chaikin2000principles}, implying that repulsive interactions between particles alone suffice to produce this structure.

Evidence for what was cautiously called `short-wavelength density fluctuations in liquid metals' \cite{Sjogren_1979} accumulated during three decades \cite{Copley_1975,DORNER19651537, Dahlborg01011977,Bermejo1997}. Only recently, however,  the term `phonon' begun to be used \cite{Khusnut2020}. Inelastic X-ray scattering experiments on neon \cite{Cusolo2001} clearly resolve a dispersion for sound, which persists above the melting temperature. 

\begin{figure}
 \centering
\includegraphics[width=0.45\textwidth]{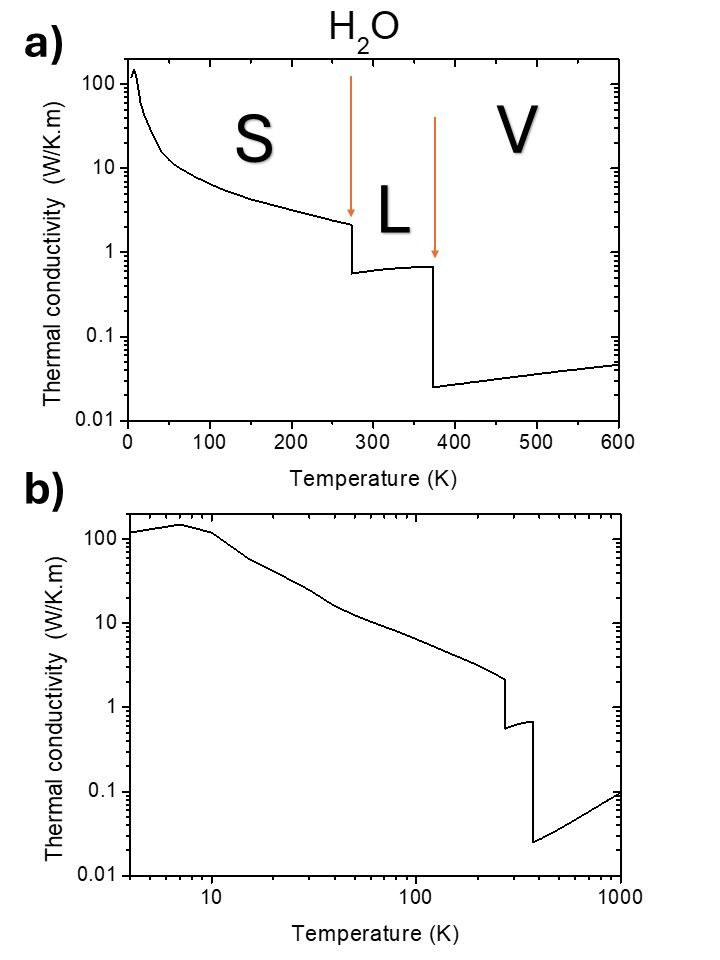}
 \caption{a) Thermal conductivity of water, ice and water vapor in a semi-logarithmic plot. b) Same data in a log-log plot.}
\label{fig2}
\end{figure}
The dynamic structure factor  of liquids has been investigated extensively by inelastic X-ray scattering, including measurements of collective density dynamics in water \cite{Monaco1999,Bencivenga2007} and in liquid metals \cite{SCOPIGNO2002121}. 

\begin{figure*}
\includegraphics[width=1\textwidth]{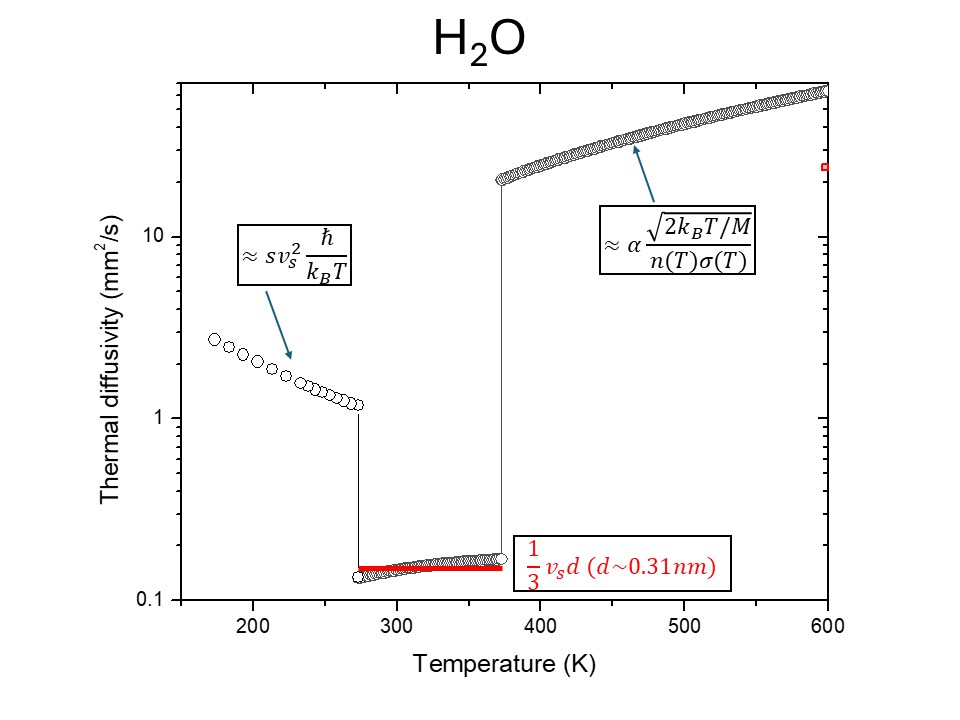}
 \caption{ Thermal diffusivity, $D$, of the three phases of H$_2$O. It has been derived from thermal conductivity, $\kappa (T)$, shown in the previous figure,  and $C_p (T)$, the specific heat per unit of mass at constant pressure, using $D=\frac{\kappa}{C_p \rho}$. Here, $\rho$ is the mass density. Expressions relevant to the three phases are also shown. In the liquid state, $d=n^{-1/3}$ is the intermolecular distance. }
\label{fig3}
\end{figure*}

\section{Thermal transport in the three phases of matter}
Figure \ref{fig2}a shows the temperature dependence of thermal conductivity, $\kappa$, of H$_2$O at ambient pressure. The thermal conductivity of ice  was reported by Slack \cite{Slack1980}. The data for liquid and vapor states are those available on the NIST website \cite{nist}. The two panels represent the same data in semi-logarithmic and logarithmic scales to make the distinction  between the three phases clear. First-order phase transitions lead to sharp drops at 273 K and at 373 K.

In the solid state, $\kappa$ has a non-monotonic  temperature dependence, displaying a peak of $\approx 120$ W.K$^{-1}$.m$^{-1}$ around 8 K. This peak separates two regimes of heat transport, a high-temperature intrinsic regime (in which thermal transport is limited by phonon-phonon collisions) and a low-temperature extrinsic regime (where thermal conductivity is governed by either disorder or boundary scattering). The intrinsic regime of heat transport in solids above the thermal conductivity peak has been extensively studied by contemporary computational condensed matter physics \cite{Tadano2018,LINDSAY2018106}. A strikingly accurate description of the amplitude and the temperature dependence of thermal conductivity has been achieved  by solving the Boltzmann-Peierls equation after computing the phonon spectrum and three-phonon scattering phase space \cite{Lindsay2013}.  However, the Ih phase of ice is proton-disordered and a microscopic calculation requires sampling over a large number of proton configurations, or the use of large supercells to approximate the configurational disorder \cite{Zivkovic2025}. 

The thermal conductivity of the liquid state  is often analyzed using molecular dynamics \cite{Puli2020}. In contrast to the solid state, its temperature dependence is monotonic and  rather weak. Finally, in the gaseous state,  thermal conductivity steadily increases with warming, a dependence much stronger than in the liquid phase.  It is remarkable that  H$_2$O, which is far from being a simple liquid, clearly displays these very different regimes of heat transport. 

Figure \ref{fig3} represents the thermal diffusivity, $D$, of H$_2$O, expressed in $mm^2/s$.  It is defined by the heat equation: 
\begin{equation}
\frac{\partial T}{\partial t}= D\nabla^2T
\label{diffusion}
\end{equation}

It is a diffusion constant for kinetic energy, and as such can be understood as the product of a characteristic velocity and a characteristic length scale. 

\begin{figure}
\includegraphics[width=0.55\textwidth]{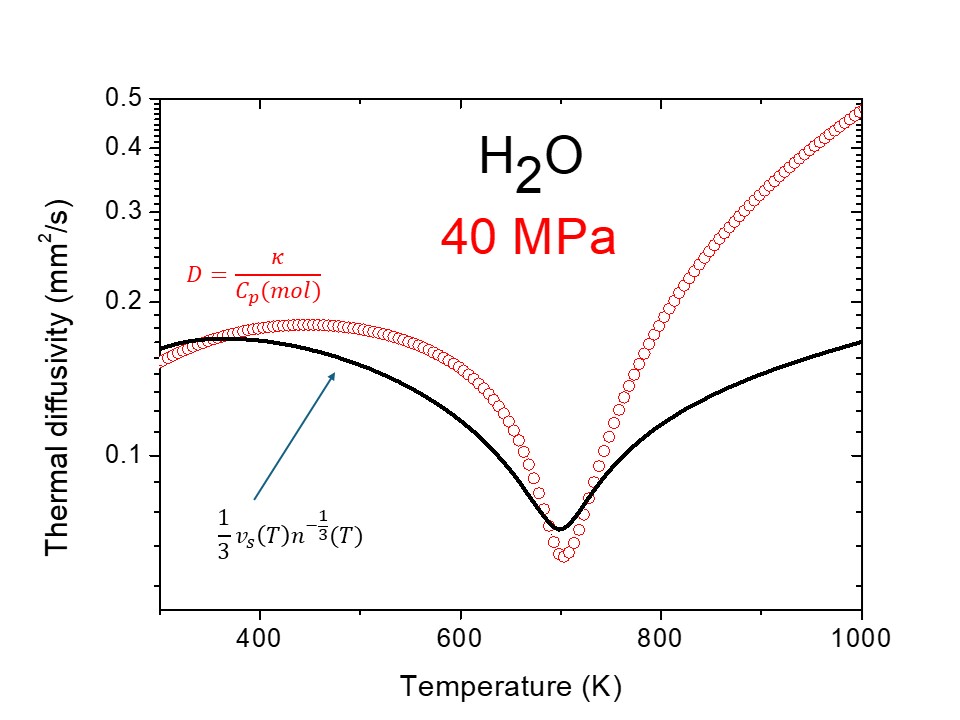}
 \caption{Supercritical H$_2$O at a pressure of 40 MPa. Even in the absence of true phase transition, its thermal diffusivity (red empty circles, extracted from thermal conductivity and specific heat at constant pressure \cite{nist}  displays a minimum. The product of intermolecular distance, $d=n^{-1/3}$,  extracted from particle density \cite{nist}, and the sound velocity \cite{nist} (black line) tracks the temperature dependence of thermal diffusivity and near the minimum, the difference  between them is less than 15 percent. }
\label{fig4}
\end{figure}

In the solid state, the relevant velocity is the speed of sound, which has little temperature dependence. The characteristic length scale is the phonon mean free path which rapidly decreases with warming. In the gaseous state, the relevant velocity is that of particles (atoms or molecules) due to their thermal agitation, which increases with warming. The relevant length scale is the particle mean-free-path, which also increases with warming. Note that in both cases, the diffusive propagation of heat is described by Equation \ref{diffusion}. But the agents of diffusion are different. In the gas, the diffusion of heat occurs thanks to collisions between bare particles. In the solid, heat diffuses thanks to collisions between phononic quasiparticles (anharmonicity). Without such collisions, there is no heat resistivity.  In the liquid state, sandwiched between them, the relevant velocity (sound) and the relevant length  scale  (inter-particle distance) both display a weak temperature dependence. As clearly seen in Figure \ref{fig3}, the temperature dependence of thermal diffusivity in water is much weaker than what is seen for ice and for water vapor.

At this stage, it would be useful to pursue a phenomenological, but quantitative, approach.

Sound velocity in solids is in the range of km/s. Trachenko and colleagues \cite{Trachenko2020} have demonstrated that there is an upper bound of $\approx $36 km/s to the sound velocity imposed by fundamental constants. The speed of sound in ice is $\approx$ 3.5 km/s \cite{Smith1986}. At the melting temperature, the mean free path of phonons in ice is $\ell_{ph}=3D (T=273 K)/v_s\approx 1$ nm, roughly twice the lattice parameter of ice. It is  instructive to scrutinize the scattering time, $\tau_{ph}$, which is the ratio of the phonon mean free path to  the sound velocity. It has been noted \cite{martelli2018,Behnia_2019} that in many solids this scattering time approaches, but remains above, the so-called Planckian dissipation time \cite{Maldacena2016,Hartnoll2022},  $\tau_{pl}=\frac{\hbar}{k_BT}$.  Mousatov and Hartnoll \cite{Mousatov2020} have  explained this feature by invoking the Lindemann criterion for melting \cite{Lawson01082009}. They deduced a correlation between sound velocity and a melting velocity, defined as $v_m=a k_BT_m/\hbar$ ($a$ is the lattice parameter, and $T_m$ is the melting temperature) \cite{Mousatov2020}. For recent case studies of the experimental relevance of this approach, see \cite{Xu2021,Martelli2021}. To sum up, in a crystalline solid, near its melting temperature, the thermal diffusivity can be written as:
\begin{equation}
D_{crys.}=s v_s^2\frac{\hbar}{k_BT}
\label{diff-solid}
\end{equation}
The numerical prefactor, $s$, is larger than unity, and is often temperature dependent. It tends to a constant  near the melting temperature. 

In a gas, the velocity is given by the thermal energy of the molecule,  $M$: $v_{th}=\sqrt{2k_BT/M}$. At 373 K, when water evaporates, the thermal velocity of gas molecules is about 500 m/s, which is almost three times lower than the sound velocity. This threefold change is overwhelmed by the change in the mean free path, which increases by more than two orders of magnitude. The mean free path of particles in a gas is roughly equal to the inverse of the product of the collision cross-section, $\sigma$, and the gas density, $n$ \cite{reif2009fundamentals}. The thermal diffusivity of a gas can be expressed as:
\begin{equation}
D_{gas}=\alpha\sqrt{\frac{2k_BT}{M}}\frac{1}{n\sigma}
\label{diff-gas}
\end{equation}

The numerical prefactor is $\alpha\simeq 0.59$ for a monoatomic gas. Subtle complications  arise in the presence of rotational and vibrational degrees of freedom \cite{THIJSSE1979307}. Let us limit ourselves to asserting that the thermal diffusivity of a gas, far larger than the two condensed states of matter, grows with temperature for two reasons. Warming enhances the average velocity and diminishes the effective collision cross-section. Faster molecules penetrate further into the steeply repulsive core of the interaction potential.

It is striking that thermal diffusivity of the liquid state appears to be the simplest. One can write:
\begin{equation}
D_{liq.}=\frac{1}{3}v_sd 
\label{diff-liq}
\end{equation}

This is the Bridgman formula, combined with the assumption that the specific heat per atom is the Boltzmann constant times a number of the order of unity. Taking the phonon mean free path to be  the distance between  water molecules, which at ambient pressure is $d\simeq$ 0.31 nm and the sound velocity of $v_s\simeq 1500 m/s$, this combination yields: $D_{liq.} \simeq 0.155 mm^2/s$. As seen in the figure, this gives a surprisingly good account of the thermal diffusivity in water, which varies between 0.14 mm$^2$/s at melting temperature to 0.17 mm$^2$/s at boiling temperature. 

Remarkably, this agreement survives under pressure when water becomes supercritical and there is no thermodynamic transition between liquid and gaseous phases.  Following an original idea put forward by Frenkel \cite{frenkel1955kinetic}, Brazhkin, Trachenko and their collaborators \cite{Brazhkin2013,Trachenko_2016} argued that the supercritical region is not featureless and a  Frenkel line on the phase diagram separates a rigid liquid from a gas-like fluid and this dynamic line has observable experimental consequences. 

Figure \ref{fig4} shows thermal diffusivity of water under a pressure of 40 MPa (400 times the ambient pressure) \cite{nist}. One can see that the
thermal diffusivity has minimum, which occurs at $\approx$ 700 K. Numerous  supercritical liquids display a similar minimum in their diffusivity \cite{Trachenko2021} .

Interestingly, in supercritical water, the amplitude of thermal diffusivity at this minimum, which is $\simeq 0.063$ mm$^2$/s, remains very close to  $D_{liq.}=\frac{1}{3}v_sd$.  At T=700 K and p=40 MPa, the sound velocity is $\simeq 500$ m/s and the  intermolecular distance is $\simeq$ 0.42 nm, so $\frac{1}{3}v_sd\simeq $0.072 mm$^2$/s, 1.14 times the observed minimum.

Given that the Bridgman formula is intimately linked to the lower bound on diffusive conduction set by the interatomic distance, it has an interesting, yet under-explored, link with the concept of the Mott-Ioffe-Regel limit \cite{Gurvitch1981,Hussey2004} in the physics of electrons in solids. 

\section{Hydrodynamics and the Bridgman formula}

To derive the formula, let us assume that in a dense liquid, energy propagates by local elastic compression waves moving at the sound velocity and then add a characteristic distance for local equilibrium  set by the liquid density.

Let us start with Euler equations for an inviscid fluid:

\begin{equation}
\frac{\partial \rho}{\partial t}
+\nabla\cdot(\rho\mathbf{v})=0,
\end{equation}

\begin{equation}
\rho
\left(
\frac{\partial\mathbf{v}}{\partial t}
+\mathbf{v}\cdot\nabla\mathbf{v}
\right)
=
-\nabla P,
\end{equation}

\begin{equation}
\rho
\frac{De}{Dt}
=
-P\nabla\cdot\mathbf{v},
\end{equation}

These equations are imposed by conservation of mass, momentum and energy.  $\rho$ is the mass density, $P$ the pressure, $e$ the internal energy per unit mass, and $\mathbf{v}$ the fluid velocity. They describe reversible adiabatic flow. Therefore, they do not contain Fourier's law and the diffusive propagation of heat. Consider a linear perturbation around equilibrium:

\begin{equation}
T=T_0+\delta T,
\end{equation}

\begin{equation}
\rho=\rho_0+\delta\rho,
\end{equation}

\begin{equation}
P=P_0+\delta P,
\end{equation}

Injecting these into Euler's equations leads us to the acoustic wave equation:

\begin{equation}
\frac{\partial^2\delta\rho}{\partial t^2}
=
v_s^2\nabla^2\delta\rho,
\end{equation}

The adiabatic sound velocity of this wave is given by:

\begin{equation}
v_s^2
=
\left(
\frac{\partial P}{\partial\rho}
\right)_S
\end{equation}

A local thermal perturbation transfers energy between neighboring regions over a characteristic distance of a=n$_{\mathrm{mol}}^{-1/3}$, where n$_{\mathrm{mol}}$ is the molecular number density. The energy is transferred over a characteristic time of $\tau=a/v_s$. 

The heat flux across a molecular interface and the temperature difference across a molecular distance define the thermal conductance. Accounting for isotropic propagation in three dimensions, the kinetic estimate of the thermal conductivity is

\begin{equation}
\kappa=\frac{1}{3}n_{\mathrm{mol}}
C_{\mathrm{mol}}k_Bv_s a.
\end{equation}

Here, $C_{\mathrm{mol}}$ is the dimensionless heat capacity per molecule. Replacing $a=n_{\mathrm{mol}}^{-1/3}$ leads to :
\begin{equation}
\kappa=\frac{1}{3}C_{\mathrm{mol}}
k_Bv_s n_{\mathrm{mol}}^{2/3}
\end{equation}

$C_{\mathrm{mol}}$ includes all contributions of the molecular degrees of freedom that are thermally active, as well as their interactions. If $f$ denotes the number of thermally active quadratic degrees of freedom per molecule, equipartition yields C$_{\mathrm{mol}}=f/2$. Therefore:

\begin{equation}
\kappa=\frac{f}{6}
k_Bv_s n_{\mathrm{mol}}^{2/3}
\end{equation}
This is equivalent to Eq. \ref{Bridg}, if $r=f/6$.  

Note that the numerical prefactor, $r$ depends on the molecular nature of the fluid. For water with three atoms in each molecule, taking $C_{\mathrm{mol}}\simeq 9$, which neglects quantum freezing of any degree of freedom, yields $r\simeq 3$. More generally, $r=f/6$ is not a universal number. 

Empirically, in polyatomic molecules $r \simeq 2$–$3$, corresponding to $f \simeq 12$–$18$, when rotational and vibrational modes are thermally activated. In contrast, in monoatomic liquefied noble gases, such as Ar or Ne, $r$ is expected to be of the order of unity. 

\begin{table*}
\begin{tabular}{cccccccc}
\hline
Liquid &T$_B$&  n& v$_s$&$\kappa_{\rm measured}$ &$\kappa_{\rm Bridg} (r=2)$ & extracted $r$ & References \\
-- &(K)&  (10$^{28}$m$^{-3}$)& (m/s)& mW/K.m & (mW/K.m)& &\\
\hline
$^3$He& 3.19&1.15 &104 &16& 14.6&2.2& \cite{Murphy1994,Huang2005}\\
$^4$He& 4.21&1.88 &180 &27& 35 &1.54 &\cite{Kerrisk1969zz}\\
\hline
\end{tabular}
\caption{The ambient-pressure boiling temperature, the density and the sound velocity of quantum liquids just below boiling. At this temperature, they can be marginally assimilated to classical liquids. The measured thermal conductivity is compared to the one extracted from Bridgman formula with $r=2$. Also, $r$ has been extracted assuming $\kappa_{\rm measured}=\kappa_{\rm Bridg}$}.
\label{Table1}
\end{table*}

\begin{figure*}
 \centering
\includegraphics[width=1\textwidth]{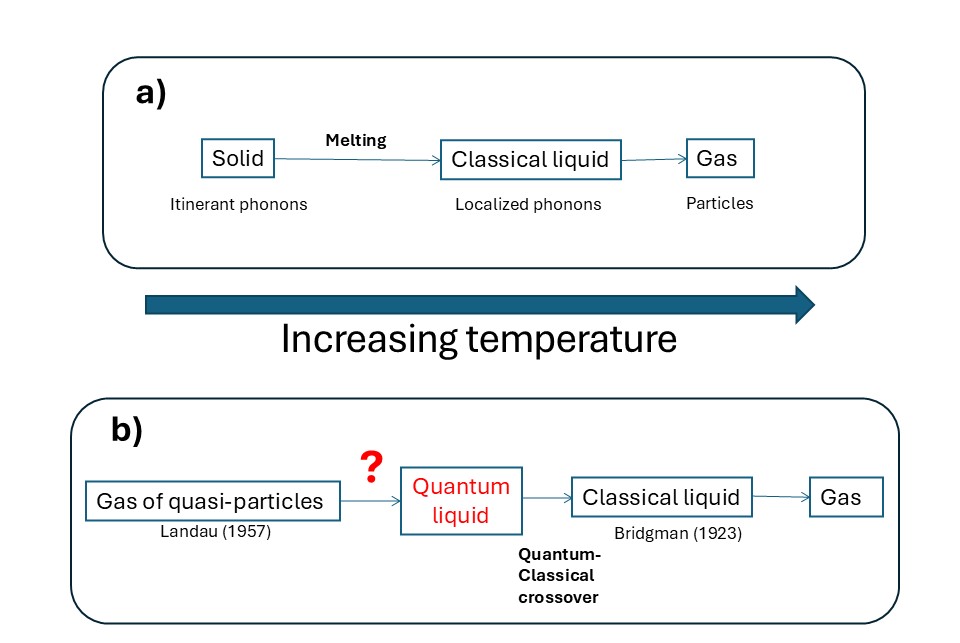}
 \caption{Sketch of different thermal regimes in a classical liquid (a) compared with a fermionic quantum liquid (b). The celebrated T$^2$ regime in Fermi liquids, corresponds to a regime in which transport is dominated by quasi-particles. Transport in the quantum liquid includes a collective component, which is yet to be fully understood. }
\label{fig5}
\end{figure*}

\section{Link to the paradox of instantaneous propagation of heat}

In the previous section, a dissipative phenomenon was derived starting from inviscid Euler equations. Entropy production was introduced via a mean free path truncation ($a=n^{-1/3}$.) It has been assumed that local thermal equilibrium is restored after an acoustic transit across a distance of $a$.  The assumption is reasonable. In water at the melting temperature, the transit time across neighboring particles ($a/v_s (liq) \simeq 0.2 ps$) and the thermal exchange time (that is the phonon lifetime, $\tau_{ph} (solid,T=273 K)\simeq 0.3 ps$) are comparable. 

The chain of arguments, which begins with Euler equations to conceive acoustic waves allowing energy transfer and ends with the introduction of an attenuation length scale is reminiscent of a paradox inherent to the heat equation publicized decades ago by  Cattaneo \cite{Cattaneo1948}. The heat equation is parabolic, according to which a localized delta-function perturbation at time $t = 0$ would generate a Gaussian response with non-zero values across the entire universe for any $t > 0$.  This appears unphysical. The resolution proposed by Cattaneo and Vernotte was the introduction of a term to the equation representing the lifetime of transient states. The heat equation, as well as the steady-state Fourier equation, remain valid approximations, but only over time scales exceeding this lifetime \cite{Vernotte1958, vernotte1958veritable,cattaneo1958sur, Joseph1989,Peluso2023}. 

At short time scales, the Fourier equation is replaced by the ``telegrapher's equation'' \cite{Goldstein1951,BAGGIOLI20201}, a hybrid between the diffusion equation and the wave equation. The Bridgman formula arises in this hybrid context. It links a diffusive quantity, $\kappa$, to a wave-like property, v$_s$. Taking the risk of over-simplification, we may qualify it as the representation of an overdamped wave-like propagation of heat. The Telegrapher's equation evolves to a diffusive equation over a time of the order of $\simeq a/v_s$. 

A similar feature has been expressed, although in a different language, by Peluso \cite{Peluso2023}, who argued that heat propagation in a liquid can be best understood by assuming a picture of dual propagation in which coexisting particles and elastic wave-packets exchange energy and momentum.  

\section{The fate of the Bridgman formula across the quantum-classical crossover}

The two isotopes of helium do not solidify under their saturated vapor pressure, even when cooled down to zero temperature. They are quantum liquids \cite{nozieres1999theory,leggett2006quantum}. What about their thermal conductivity? 

Table \ref{Table1} lists their properties at the boiling temperature. Their density and sound velocity injected to Eq. \ref{Bridg}, and assuming r=2, yields a Bridgman thermal conductivity, comparable to the experimentally measured value.  If one assumes the validity of the Bridgman formula, their measured thermal conductivity just below the boiling temperature $r\simeq$ 1.8 for $^4$He  and $r\simeq 2.2$ for $^3$He. These $r$ values are larger than $r=1$ naively expected given their monoatomic nature.  
The degeneracy temperature for bosons and fermions of their density is only slightly lower than the boiling temperature and the relevant quantum statistics of each can no longer be neglected.  
The bosonic $^4$He is superfluid below 2.17 K, while the fermionic $^3$He does not go through any phase transition down to $\approx$ 2 mK. In these conditions, a discrepancy between $\kappa_{measured}$ and $\kappa_{Bridg}$ emerges and widens on cooling.

Over a wide range (by three orders of magnitude) of temperature, $^3$He is a liquid of fermionic quasi-particles \cite{Dobbs}, which inspired Landau's Fermi liquid theory \cite{Landau1957} in the 1950s. The first theoretical study of transport properties of a Fermi liquid \cite{Abrikosov_1959} found that at low temperature, the thermal conductivity should be proportional to the inverse of temperature. Soon this was confirmed by other theoretical studies \cite{nozieres1999theory,Dy1969,wolfle1979} and a string of experimental studies \cite{Anderson1966,abel1967,Wheatley,greywall1984, Mitchell1986, Murphy1994} carried out in the following decades (see \cite{Huang2012} for a compilation of experimental data).

The expected $\kappa \propto$ T$^{-1}$ can be traced back to the combination of a T-linear specific heat  (C$_V \propto$ T) with thermal diffusivity (D$ \propto$ T$^{-2}$). The latter is a consequence of the thermal evolution of the fermion-fermion scattering phase space ($\propto$ T$^2$/T$^2_F$) \cite{KADOWAKI,Hussey2005,Behnia2022}. This is a particle-based picture of transport, which neglects collective transport.

For several decades, a striking experimental fact in $^3$He was overlooked: When temperature exceeds $\approx 0.1$T$_F$, the temperature dependence of the experimentally measured thermal conductivity becomes $\kappa \propto$ T$^{1/2}$, in sharp contrast to what was theoretically expected. During these decades,  the  experimentally measured thermal conductivity of liquid $^3$He at the boiling temperature was not compared to what is expected by the Bridgman formula, which was never invoked as a point of reference.  

In 2024, the present author and Kostya Trachenko \cite{Behnia2024}, proposed an explanation for the $\propto$ T$^{1/2}$ temperature dependence of thermal conductivity in normal liquid $^3$He. 

This proposal can be considered as a quantum version of the Bridgman formula. It led to equation 4 in \cite{Behnia2024}) in which the collective (sound-based) contribution to thermal conductivity is:

\begin{equation}
 \kappa_s=r k_Bn^{2/3}\sqrt{\frac{2k_BT}{m^\star}}
 \label{BK}
\end{equation}

Here, $r=\pi^{11/6}/3^{4/3}\approx 1.88$ and $m^\star$ is the effective mass. Eq.\ref{BK} is Eq. \ref{Bridg}, with sound velocity replaced by thermal velocity and with a slightly modified numerical prefactor. 

Its derivation \cite{Behnia2024} was inspired by Landauer's picture of conduction as transmission \cite{Imry1999}. It began by assuming that each fermionic state  carries a quantum of thermal conductance, $\pi^2/3 \frac{k_B^2 T}{h}$ over a distance of the order of quasi-particle mean-free-path. The latter was set by scattering rate of quasi-particles. Counting the total number of available states led to a contribution by quasi-particles in agreement with the established temperature dependence ($\kappa_{qp}\propto T^{-1}$) and amplitude. 

Then, it was assumed that in addition to this particle-based channel, there is a second one, $\kappa_s$, in which the whole Fermi surface  carries a quantum of thermal conductance over a distance as long as the de Broglie thermal length $\Lambda =\frac{h}{\sqrt{2\pi m^\star k_BT}}$. This led to Eq. \ref{BK} for $\kappa_s$. The effective mass is $m^\star\simeq2.7m_3$, where $m_3$ is the bare mass of $^3$He. The reduced Planck constant,$\hbar$, due to a cancellation between its presence in both $\Lambda$ and in $\frac{k_B^2}{h}$, is absent in Eq.\ref{BK}.

In the proposed scenario \cite{Behnia2024} heat is carried by both quasi-particles ($\kappa_{qp}\propto$ T$^{-1}$) and sound ($\kappa_{s}\propto$ T$^{1/2}$). The  latter smoothly transforms to the Bridgman formula across the quantum-classical crossover. This phenomenological picture of two-channel transport could give a quantitative account of the experimentally measured thermal conductivity of $^3$He \cite{greywall1984,Murphy1994} between $0.02 K<$T$<3$ K. The disagreement between the experimental data and what is expected by Equation \ref{BK} is less than twenty percent.

Note that the phenomenological success of Eq. \ref{BK} in explaining the large discrepancy between the experimental data and the particle-based picture of transport in $^3$He is not sufficient to prove that most of heat is carried by sound. However, it is fair to say that at present it is the most plausible available explanation of what happens between 0.02K (below which the particle-based picture works) and 3.2K (at which the classical Bridgman formula holds) (see figure \ref{fig5}). 

An open question is the relevance of such a collective mode of transport to metallic Fermi liquids \cite{Behnia2026}. 

 This work was supported by the Agence Nationale de la Recherche (ANR-25-CE30-5730-01). The author is grateful to University of California, Berkeley and to James Analytis for hosting him as an ICAM fellow, which allowed this paper to be written. The stay was funded by the US National Science Foundation (NSF) Grant Number 2201516 under the Accelnet program of the Office of International Science and Engineering (OISE).

\bibliography{biblio}

@book{ziman1972,
  title={Principles of the Theory of Solids},
  author={Ziman, J.M.},
  year={1972},
  publisher={Cambridge University Press}

}

@BOOK{Dobbs,
   author = {Dobbs, E R},
    title = "{Helium Three}",
     year = 2000,
     publisher = {Oxford University Press}
}

@article{KADOWAKI,
title = {Universal relationship of the resistivity and specific heat in heavy-Fermion compounds},
journal = {Solid State Communications},
volume = {58},
number = {8},
pages = {507-509},
year = {1986},
issn = {0038-1098},
author = {K. Kadowaki and S.B. Woods}
}

@article{Abrikosov_1959,
	year = {1959},
	publisher = {{IOP} Publishing},
	volume = {22},
	number = {1},
	pages = {329-367},
	author = {A A Abrikosov and I M Khalatnikov},
	title = {The theory of a {F}ermi liquid (the properties of liquid {$^3$He} at low temperatures)},
	journal = {Reports on Progress in Physics},
}

@article{wolfle1979,
	year = 1979,
	publisher = {{IOP} Publishing},
	volume = {42},
	pages = {269-346},
	author = {P Wolfle},
	title = {Low-temperature properties of liquid {$^3$H}e},
	journal = {Reports on Progress in Physics},
}

@article{greywall1984,
  title={Thermal conductivity of normal liquid {$^3$He}},
  author={Greywall, Dennis S},
  journal={Physical Review B},
  volume={29},
  number={9},
  pages={4933},
  year={1984},
  publisher={APS}}

@article{Wheatley,
  title = {Experimental properties of liquid ${\mathrm{He}}^{3}$ near the absolute zero},
  author = {Wheatley, J. C.},
  journal = {Physical Review},
  volume = {165},
pages = {304-309},
  year = {1968},
  publisher = {American Physical Society}}

@article{abel1967,
  title = {Thermal conductivity of pure {${\mathrm{He}}^{3}$} and of dilute solutions of {${\mathrm{He}}^{3}$} in {${\mathrm{He}}^{4}$} at low temperatures},
  author = {Abel, W. R. and Johnson, R. T. and Wheatley, J. C. and Zimmermann, W.},
  journal = {Physical Review Letters},
  volume = {18},
  pages = {737-740},
  year = {1967}
  }

@article{martelli2018,
  title = {Thermal Transport and Phonon Hydrodynamics in Strontium Titanate},
  author = {Martelli, Valentina and Jim\'enez, Julio Larrea and Continentino, Mucio and Baggio-Saitovitch, Elisa and Behnia, Kamran},
  journal = {Physical Review Letters},
  volume = {120},
  pages = {125901},
  numpages = {6},
  year = {2018},
  month = {Mar},
  publisher = {American Physical Society},
  doi = {10.1103/PhysRevLett.120.125901},
  url = {https://link.aps.org/doi/10.1103/PhysRevLett.120.125901}
}

@article{Hussey2005,
author = {E. Hussey ,N.},
title = {Non-generality of the {Kadowaki–Woods} Ratio in Correlated Oxides},
journal = {Journal of the Physical Society of Japan},
volume = {74},
number = {4},
pages = {1107-1110},
year = {2005},
doi = {10.1143/JPSJ.74.1107}
}

@article{Landau1957,
author = {Landau, Lev},
title = {{The theory of a Fermi liquid}},
  journal = {Soviet Journal of Experimental and Theoretical Physics},
     year = 1957,
   volume = 3,
    pages = {921},
}

@article{Dy1969,
  title = {Transport Coefficients of a Normal Fermi Liquid: Application to Liquid ${\mathrm{He}}^{3}$},
  author = {Dy, K. S. and Pethick, C. J.},
  journal = {Phys. Rev.},
  volume = {185},
  issue = {1},
  pages = {373--384},
  numpages = {0},
  year = {1969},
  month = {Sep},
  publisher = {American Physical Society},
  doi = {10.1103/PhysRev.185.373},
  url = {https://link.aps.org/doi/10.1103/PhysRev.185.373}
}

@Article{Behnia2024,
author={Behnia, Kamran
and Trachenko, Kostya},
title={How heat propagates in liquid {$^3$He}},
journal={Nature Communications},
year={2024},
month={Feb},
day={27},
volume={15},
number={1},
pages={1771},
issn={2041-1723},
doi={10.1038/s41467-024-46079-0},
url={https://doi.org/10.1038/s41467-024-46079-0}
}

@article{Behnia2022,
author = {Behnia, Kamran},
title = {On the Origin and the Amplitude of {T}-Square Resistivity in Fermi Liquids},
journal = {Annalen der Physik},
volume = {534},
number = {5},
pages = {2100588},
doi = {https://doi.org/10.1002/andp.202100588},
url = {https://onlinelibrary.wiley.com/doi/abs/10.1002/andp.202100588},
year = {2022}
}

@Article{Murphy1994,
author={Murphy, Daniel
and Meyer, Horst},
title={The thermal conductivity of and convective onset in liquid {3He} at saturated vapor pressure above 1.3 K},
journal={Journal of Low Temperature Physics},
year={1994},
month={Dec},
day={01},
volume={97},
number={5},
pages={509-517},
issn={1573-7357},
doi={10.1007/BF00754305},
url={https://doi.org/10.1007/BF00754305}
}

@article{Brigman1923,
 ISSN = {01999818},
 URL = {http://www.jstor.org/stable/20026073},
 author = {P. W. Bridgman},
 journal = {Proceedings of the American Academy of Arts and Sciences},
 number = {7},
 pages = {141--169},
 publisher = {American Academy of Arts & Sciences},
 title = {The Thermal Conductivity of Liquids under Pressure},
 urldate = {2023-06-27},
 volume = {59},
 year = {1923}
}

@article{KHRAPAK2023121786,
title = {{Bridgman} formula for the thermal conductivity of atomic and molecular liquids},
journal = {Journal of Molecular Liquids},
volume = {381},
pages = {121786},
year = {2023},
issn = {0167-7322},
doi = {https://doi.org/10.1016/j.molliq.2023.121786},
url = {https://www.sciencedirect.com/science/article/pii/S0167732223005895},
author = {S.A. Khrapak}
}

@article{Ohtori2014,
  title = {Thermal conductivity of simple liquids: Temperature and packing-fraction dependence},
  author = {Ohtori, Norikazu and Ishii, Yoshiki and Togawa, Yoshinori and Oono, Takuya and Takase, Keiichi},
  journal = {Phys. Rev. E},
  volume = {89},
  issue = {2},
  pages = {022129},
  numpages = {5},
  year = {2014},
  month = {Feb},
  publisher = {American Physical Society},
  doi = {10.1103/PhysRevE.89.022129},
  url = {https://link.aps.org/doi/10.1103/PhysRevE.89.022129}
}

@article{Trachenko_2016,
doi = {10.1088/0034-4885/79/1/016502},
url = {https://doi.org/10.1088/0034-4885/79/1/016502},
year = {2015},
month = {dec},
publisher = {IOP Publishing},
volume = {79},
number = {1},
pages = {016502},
author = {Trachenko, K and Brazhkin, V V},
title = {Collective modes and thermodynamics of the liquid state},
journal = {Reports on Progress in Physics}
}

@book{trachenko2023theory,
  title={Theory of Liquids: From Excitations to Thermodynamics},
  author={Trachenko, K.},
  isbn={9781009355476},
  lccn={2023028307},
  url={https://books.google.fr/books?id=8d_LEAAAQBAJ},
  year={2023},
  publisher={Cambridge University Press}
}

@article{Zhao2021,
    author = {Zhao, Andrew Z. and Wingert, Matthew C. and Chen, Renkun and Garay, Javier E.},
    title = {Phonon gas model for thermal conductivity of dense, strongly interacting liquids},
    journal = {Journal of Applied Physics},
    volume = {129},
    number = {23},
    pages = {235101},
    year = {2021},
    month = {06},
    issn = {0021-8979},
    doi = {10.1063/5.0040734},
    url = {https://doi.org/10.1063/5.0040734},
}

@book{chaikin2000principles,
  title={Principles of Condensed Matter Physics},
  author={Chaikin, P.M. and Lubensky, T.C.},
  isbn={9780521794503},
  lccn={93044244},
  url={https://books.google.fr/books?id=P9YjNjzr9OIC},
  year={2000},
  publisher={Cambridge University Press}
}

@book{de1993physics,
  title={The Physics of Liquid Crystals},
  author={de Gennes, P.G. and Prost, J.},
  isbn={9780198517856},
  lccn={93009154},
  series={International Series of Monographs on Physics},
  url={https://books.google.fr/books?id=0Nw-dzWz5agC},
  year={1993},
  publisher={Clarendon Press}
}

@article{Yarnell1973,
  title = {Structure Factor and Radial Distribution Function for Liquid Argon at 85 \ifmmode^\circ\else\textdegree\fi{}K},
  author = {Yarnell, J. L. and Katz, M. J. and Wenzel, R. G. and Koenig, S. H.},
  journal = {Phys. Rev. A},
  volume = {7},
  issue = {6},
  pages = {2130--2144},
  numpages = {0},
  year = {1973},
  month = {Jun},
  publisher = {American Physical Society},
  doi = {10.1103/PhysRevA.7.2130},
  url = {https://link.aps.org/doi/10.1103/PhysRevA.7.2130}
}

@article{Viswanath,
author = {Viswanath, D. S.},
title = {On thermal conductivity of liquids},
journal = {AIChE Journal},
volume = {13},
number = {5},
pages = {850-853},
doi = {https://doi.org/10.1002/aic.690130508},
url = {https://aiche.onlinelibrary.wiley.com/doi/abs/10.1002/aic.690130508},
year = {1967}
}

@article{Rao,
  title = {Thermal Conductivity of Liquids},
  author = {Rao, M. Rama},
  journal = {Phys. Rev.},
  volume = {59},
  issue = {2},
  pages = {212},
  numpages = {0},
  year = {1941},
  month = {Jan},
  publisher = {American Physical Society},
  doi = {10.1103/PhysRev.59.212},
  url = {https://link.aps.org/doi/10.1103/PhysRev.59.212}
}

@article{Khrapak2021,
  title = {Vibrational model of thermal conduction for fluids with soft interactions},
  author = {Khrapak, Sergey A.},
  journal = {Phys. Rev. E},
  volume = {103},
  issue = {1},
  pages = {013207},
  numpages = {7},
  year = {2021},
  month = {Jan},
  publisher = {American Physical Society},
  doi = {10.1103/PhysRevE.103.013207},
  url = {https://link.aps.org/doi/10.1103/PhysRevE.103.013207}
}

@Article{Khrapak2024,
author={Khrapak, S. A.
and Khrapak, A. G.},
title={Modified {Bridgman} Formula for the Thermal Conductivity of Complex (Dusty) Plasma Fluids},
journal={JETP Letters},
year={2024},
month={Aug},
day={01},
volume={120},
number={4},
pages={236-240},
issn={1090-6487},
doi={10.1134/S0021364024602628},
url={https://doi.org/10.1134/S0021364024602628}
}

@Article{Qing2020,
title = {A Ubiquitous Thermal Conductivity Formula for Liquids, Polymer Glass, and Amorphous Solids},
journal = {Chin. Phys. Lett.},
volume = {37},
number = {10},
pages = {104401-104401},
year = {2020},
issn = {},
doi = {10.1088/0256-307X/37/10/104401},	
url = {http://cpl.iphy.ac.cn/en/article/doi/10.1088/0256-307X/37/10/104401},
author = {Qing Xi and Jinxin Zhong and Jixiong He and Xiangfan Xu and Tsuneyoshi Nakayama and Yuanyuan Wang and Jun Liu and Jun Zhou and Baowen Li}
}

@Article{Peluso2023,
AUTHOR = {Peluso, Fabio},
TITLE = {How Does Heat Propagate in Liquids?},
JOURNAL = {Liquids},
VOLUME = {3},
YEAR = {2023},
NUMBER = {1},
PAGES = {92--117},
URL = {https://www.mdpi.com/2673-8015/3/1/9},
ISSN = {2673-8015},
DOI = {10.3390/liquids3010009}
}

@Article{Bolmatov2012,
author={Bolmatov, D.
and Brazhkin, V. V.
and Trachenko, K.},
title={The phonon theory of liquid thermodynamics},
journal={Scientific Reports},
year={2012},
month={May},
day={24},
volume={2},
number={1},
pages={421},
issn={2045-2322},
doi={10.1038/srep00421},
url={https://doi.org/10.1038/srep00421}
}

@book{bird2006transport,
  title={Transport Phenomena},
  author={Bird, R.B. and Stewart, W.E. and Lightfoot, E.N.},
  number={vol.~1},
  isbn={9780470115398},
  lccn={2001023739},
  series={Transport Phenomena},
  year={2006},
  publisher={Wiley}
}

@article{Hubbard_1969,
doi = {10.1088/0022-3719/2/3/318},
url = {https://doi.org/10.1088/0022-3719/2/3/318},
year = {1969},
month = {mar},
publisher = {},
volume = {2},
number = {3},
pages = {556},
author = {J Hubbard and J L Beeby},
title = {Collective motion in liquids},
journal = {Journal of Physics C: Solid State Physics}
}

@article{Cusolo2001,
    author = {Cunsolo, A. and Pratesi, G. and Verbeni, R. and Colognesi, D. and Masciovecchio, C. and Monaco, G. and Ruocco, G. and Sette, F.},
    title = {Microscopic relaxation in supercritical and liquid neon},
    journal = {The Journal of Chemical Physics},
    volume = {114},
    number = {5},
    pages = {2259-2267},
    year = {2001},
    month = {02},
    issn = {0021-9606},
    doi = {10.1063/1.1334613},
    url = {https://doi.org/10.1063/1.1334613},
}

@article{Sjogren_1979,
doi = {10.1088/0022-3719/12/3/011},
url = {https://doi.org/10.1088/0022-3719/12/3/011},
year = {1979},
month = {feb},
publisher = {},
volume = {12},
number = {3},
pages = {425},
author = {L Sjogren},
title = {Short-wavelength density fluctuations in liquid metals},
journal = {Journal of Physics C: Solid State Physics}
}

@article{Copley_1975,
doi = {10.1088/0034-4885/38/4/001},
url = {https://doi.org/10.1088/0034-4885/38/4/001},
year = {1975},
month = {apr},
publisher = {},
volume = {38},
number = {4},
pages = {461},
author = {J R D Copley and S W Lovesey},
title = {The dynamic properties of monatomic liquids},
journal = {Reports on Progress in Physics}
}

@article{DORNER19651537,
title = {High frequency waves in liquid lead},
journal = {Physica},
volume = {31},
number = {10},
pages = {1537-1539},
year = {1965},
issn = {0031-8914},
doi = {https://doi.org/10.1016/0031-8914(65)90120-5},
url = {https://www.sciencedirect.com/science/article/pii/0031891465901205},
author = {B. Dorner and T. Plesser and H. Stiller}
}

@article{Khusnut2020,
  title = {Collective modes and gapped momentum states in liquid {Ga}: Experiment, theory, and simulation},
  author = {Khusnutdinoff, R. M. and Cockrell, C. and Dicks, O. A. and Jensen, A. C. S. and Le, M. D. and Wang, L. and Dove, M. T. and Mokshin, A. V. and Brazhkin, V. V. and Trachenko, K.},
  journal = {Phys. Rev. B},
  volume = {101},
  issue = {21},
  pages = {214312},
  numpages = {9},
  year = {2020},
  month = {Jun},
  publisher = {American Physical Society},
  doi = {10.1103/PhysRevB.101.214312},
  url = {https://link.aps.org/doi/10.1103/PhysRevB.101.214312}
}

@article{Bermejo1997,
  title = {Collective, short-wavelength excitations in liquid gallium},
  author = {Bermejo, F. J. and Fern\'andez-Perea, R. and Alvarez, M. and Roessli, B. and Fischer, H. E. and Bossy, J.},
  journal = {Phys. Rev. E},
  volume = {56},
  issue = {3},
  pages = {3358--3369},
  numpages = {0},
  year = {1997},
  month = {Sep},
  publisher = {American Physical Society},
  doi = {10.1103/PhysRevE.56.3358},
  url = {https://link.aps.org/doi/10.1103/PhysRevE.56.3358}
}

@article{Dahlborg01011977,
author = {U. Dahlborg and M. Davidovic and K. E. Larsson},
title = {The structure factor for liquid lead},
journal = {Physics and Chemistry of Liquids},
volume = {6},
number = {3},
pages = {149--166},
year = {1977},
publisher = {Taylor \& Francis},
doi = {10.1080/00319107708084137},
URL = {https://doi.org/10.1080/00319107708084137}
}

@article{nist,
  journal={National Institute of Standards and Technology database, https://webbook.nist.gov/chemistry/fluid},
  Author = {NIST},
  year = {2023}
}

@article{Slack1980,
  title = {Thermal conductivity of ice},
  author = {Slack, Glen A.},
  journal = {Phys. Rev. B},
  volume = {22},
  issue = {6},
  pages = {3065--3071},
  numpages = {0},
  year = {1980},
  month = {Sep},
  publisher = {American Physical Society},
  doi = {10.1103/PhysRevB.22.3065},
  url = {https://link.aps.org/doi/10.1103/PhysRevB.22.3065}
}

@article{Tadano2018,
author = {Tadano ,Terumasa and Tsuneyuki ,Shinji},
title = {First-Principles Lattice Dynamics Method for Strongly Anharmonic Crystals},
journal = {Journal of the Physical Society of Japan},
volume = {87},
number = {4},
pages = {041015},
year = {2018},
doi = {10.7566/JPSJ.87.041015}
}

@article{LINDSAY2018106,
title = {Survey of ab initio phonon thermal transport},
journal = {Materials Today Physics},
volume = {7},
pages = {106-120},
year = {2018},
issn = {2542-5293},
doi = {https://doi.org/10.1016/j.mtphys.2018.11.008},
url = {https://www.sciencedirect.com/science/article/pii/S254252931830141X},
author = {L. Lindsay and C. Hua and X.L. Ruan and S. Lee}
}

@article{Lindsay2013,
  title = {Ab initio thermal transport in compound semiconductors},
  author = {Lindsay, L. and Broido, D. A. and Reinecke, T. L.},
  journal = {Phys. Rev. B},
  volume = {87},
  issue = {16},
  pages = {165201},
  numpages = {15},
  year = {2013},
  month = {Apr},
  publisher = {American Physical Society},
  doi = {10.1103/PhysRevB.87.165201},
  url = {https://link.aps.org/doi/10.1103/PhysRevB.87.165201}
}

@article{Zivkovic2025,
    author = {Zivkovic, Aleksandar and Terranova, Umberto and de Leeuw, Nora H.},
    title = {Water Is Cool: Advanced Phonon Dynamics in Ice Ih and Ice XI via Machine Learning Potentials and Quantum Nuclear Vibrations},
    journal = {Journal of Chemical Theory and Computation},
    volume = {21},
    number = {4},
    pages = {1978-1989},
    year = {2025},
    month = {02},
    issn = {1549-9618},
    doi = {10.1021/acs.jctc.4c01582},
}

@article{Puli2020,
  title = {Atomistic simulations of the thermal conductivity of liquids},
  author = {Puligheddu, Marcello and Galli, Giulia},
  journal = {Phys. Rev. Mater.},
  volume = {4},
  issue = {5},
  pages = {053801},
  numpages = {6},
  year = {2020},
  month = {May},
  publisher = {American Physical Society},
  doi = {10.1103/PhysRevMaterials.4.053801},
  url = {https://link.aps.org/doi/10.1103/PhysRevMaterials.4.053801}
}

@article{THIJSSE1979307,
title = {Some simplified expressions for the thermal conductivity in an external field},
journal = {Physica A: Statistical Mechanics and its Applications},
volume = {98},
number = {1},
pages = {307-312},
year = {1979},
issn = {0378-4371},
doi = {https://doi.org/10.1016/0378-4371(79)90181-X},
url = {https://www.sciencedirect.com/science/article/pii/037843717990181X},
author = {B.J. Thijsse and G.W. {'T Hooft} and D.A. Coombe and H.F.P. Knaap and J.J.M. Beenakker}
}

@article{Trachenko2020,
author = {K. Trachenko  and B. Monserrat  and C. J. Pickard  and V. V. Brazhkin },
title = {Speed of sound from fundamental physical constants},
journal = {Science Advances},
volume = {6},
number = {41},
pages = {eabc8662},
year = {2020},
doi = {10.1126/sciadv.abc8662}}

@article{Smith1986,
author = {Smith, Alphonso C. and Kishoni, Doron},
title = {Measurement of the speed of sound in ice},
journal = {AIAA Journal},
volume = {24},
number = {10},
pages = {1713-1715},
year = {1986},
doi = {10.2514/3.9510}
}

@article{Behnia_2019,
  doi = {10.1088/1361-648x/ab2db6},
  url = {https://doi.org/10.1088%2F1361-648x%2Fab2db6},
  year = {2019},
  month = {jul},
  publisher = {{IOP} Publishing},
  volume = {31},
  number = {40},
  pages = {405702},
  author = {Kamran Behnia and Aharon Kapitulnik},
  title = {A lower bound to the thermal diffusivity of insulators},
  journal = {J. of Phys.: Condens. Matter},
}

@article{Mousatov2020,
  title = {On the {Planckian} bound for heat diffusion in insulators},
  author = {Mousatov, Connie H and Hartnoll, Sean A},
  journal = {Nat. Phys.},
  volume = {16},
  pages = {579–584},
  year = {2020},
  doi = {10.1038/s41567-020-0828-6},
  publisher = {Nature Publishing Group},
}

@article{Xu2021,
  title = {Thermal conductivity of bulk {${\mathrm{In}}_{2}{\mathrm{O}}_{3}$} single crystals},
  author = {Xu, Liangcai and Fauqu\'e, Beno\^{\i}t and Zhu, Zengwei and Galazka, Zbigniew and Irmscher, Klaus and Behnia, Kamran},
  journal = {Phys. Rev. Mater.},
  volume = {5},
  issue = {1},
  pages = {014603},
  numpages = {7},
  year = {2021},
  month = {Jan},
  publisher = {American Physical Society},
  doi = {10.1103/PhysRevMaterials.5.014603},
  url = {https://link.aps.org/doi/10.1103/PhysRevMaterials.5.014603}
}

@article{Hartnoll2022,
  title = {Colloquium: Planckian dissipation in metals},
  author = {Hartnoll, Sean A. and Mackenzie, Andrew P.},
  journal = {Rev. Mod. Phys.},
  volume = {94},
  issue = {4},
  pages = {041002},
  numpages = {27},
  year = {2022},
  month = {Nov},
  publisher = {American Physical Society},
  doi = {10.1103/RevModPhys.94.041002},
  url = {https://link.aps.org/doi/10.1103/RevModPhys.94.041002}
}

@article{Martelli2021,
  title = {Thermal diffusivity and its lower bound in orthorhombic {SnSe}},
  author = {Martelli, Valentina and Abud, Fabio and Jim\'enez, Julio Larrea and Baggio-Saitovich, Elisa and Zhao, Li-Dong and Behnia, Kamran},
  journal = {Phys. Rev. B},
  volume = {104},
  issue = {3},
  pages = {035208},
  numpages = {7},
  year = {2021},
  month = {Jul},
  publisher = {American Physical Society},
  doi = {10.1103/PhysRevB.104.035208},
  url = {https://link.aps.org/doi/10.1103/PhysRevB.104.035208}
}

@Article{Maldacena2016,
author={Maldacena, Juan
and Shenker, Stephen H.
and Stanford, Douglas},
title={A bound on chaos},
journal={Journal of High Energy Physics},
year={2016},
month={Aug},
day={17},
volume={2016},
number={8},
pages={106},
issn={1029-8479},
doi={10.1007/JHEP08(2016)106},
url={https://doi.org/10.1007/JHEP08(2016)106}
}

@article{Lawson01082009,
author = {A.C. Lawson},
title = {Physics of the {Lindemann} melting rule},
journal = {Philosophical Magazine},
volume = {89},
number = {22-24},
pages = {1757--1770},
year = {2009},
publisher = {Taylor \& Francis},
doi = {10.1080/14786430802577916}}

@book{reif2009fundamentals,
  title={Fundamentals of Statistical and Thermal Physics},
  author={Reif, F.},
  isbn={9781478610052},
  year={2009},
  publisher={Waveland Press}
}

@article{Hussey2004,
author = {N. E. Hussey and K. Takenaka and H. Takagi},
title = {Universality of the {Mott–Ioffe–Regel} limit in metals},
journal = {Philosophical Magazine},
volume = {84},
number = {27},
pages = {2847--2864},
year = {2004},
publisher = {Taylor \& Francis},
doi = {10.1080/14786430410001716944}}

@article{Gurvitch1981,
  title = {{Ioffe-Regel} criterion and resistivity of metals},
  author = {Gurvitch, M.},
  journal = {Phys. Rev. B},
  volume = {24},
  issue = {12},
  pages = {7404--7407},
  numpages = {0},
  year = {1981},
  month = {Dec},
  publisher = {American Physical Society},
  doi = {10.1103/PhysRevB.24.7404},
  url = {https://link.aps.org/doi/10.1103/PhysRevB.24.7404}
}

@article{Brazhkin2013,
  title = {``Liquid-Gas'' Transition in the Supercritical Region: Fundamental Changes in the Particle Dynamics},
  author = {Brazhkin, V. V. and Fomin, Yu. D. and Lyapin, A. G. and Ryzhov, V. N. and Tsiok, E. N. and Trachenko, Kostya},
  journal = {Phys. Rev. Lett.},
  volume = {111},
  issue = {14},
  pages = {145901},
  numpages = {5},
  year = {2013},
  month = {Oct},
  publisher = {American Physical Society},
  doi = {10.1103/PhysRevLett.111.145901},
  url = {https://link.aps.org/doi/10.1103/PhysRevLett.111.145901}
}

@article{Trachenko2021,
  title = {Universal lower bounds on energy and momentum diffusion in liquids},
  author = {Trachenko, K. and Baggioli, M. and Behnia, K. and Brazhkin, V. V.},
  journal = {Phys. Rev. B},
  volume = {103},
  issue = {1},
  pages = {014311},
  numpages = {7},
  year = {2021},
  month = {Jan},
  publisher = {American Physical Society},
  doi = {10.1103/PhysRevB.103.014311},
  url = {https://link.aps.org/doi/10.1103/PhysRevB.103.014311}
}

@book{frenkel1955kinetic,
  title={Kinetic Theory of Liquids},
  author={Frenkel, I.A.I.},
  lccn={56013521},
  series={Dover books on chemistry and physical chemistry},
  year={1955},
  publisher={Dover Publications}
}

@article{Kerrisk1969zz,
    author = "Kerrisk, Jerry F. and Keller, William E.",
    title = "{Thermal Conductivity of Fluid He-3 and He-4 at Temperatures between 1.5 and 4.0-degrees K and for Pressures up to 34 atm}",
    doi = "10.1103/PhysRev.177.341",
    journal = "Phys. Rev.",
    volume = "177",
    pages = "341--351",
    year = "1969"
}

@article{Cattaneo1948,
  author  = {Cattaneo, Carlo},
  title   = {Sulla conduzione del calore},
  journal = {Atti del Seminario Matematico e Fisico dell'Universit{\`a} di Modena},
  volume  = {3},
  pages   = {83--101},
  year    = {1948}
}

@article{Joseph1989,
  title = {Heat waves},
  author = {Joseph, D. D. and Preziosi, Luigi},
  journal = {Rev. Mod. Phys.},
  volume = {61},
  issue = {1},
  pages = {41--73},
  numpages = {0},
  year = {1989},
  month = {Jan},
  publisher = {American Physical Society},
  doi = {10.1103/RevModPhys.61.41},
  url = {https://link.aps.org/doi/10.1103/RevModPhys.61.41}
}

@article{Vernotte1958,
author="Vernotte, P.",
title="Les paradoxes de la theorie continue de l'\'equation de la chaleur",
journal="Comptes Rendus",
year="1958",
volume="246",
pages="3154",
URL="https://cir.nii.ac.jp/crid/1570572700876716288"
}

@article{vernotte1958veritable,
  title={La veritable \'equation de la chaleur},
  author={Vernotte, P},
  journal={CR Acad. Sci},
  volume={247},
  pages={2103--2105},
  year={1958}
}

@article{cattaneo1958sur,
  author          = {Cattaneo, Carlo},
  title           = {Sur une forme de l'{\'e}quation de la chaleur {\'e}liminant le paradoxe d'une propagation instantan{\'e}e},
  journal         = {Comptes Rendus Hebdomadaires des S{\'e}ances de l'Acad{\'e}mie des Sciences},
  volume          = {247},
  pages           = {431--433},
  year            = {1958}
}

@article{Goldstein1951,
    author = {Goldstein, S.},
    title = {ON DIFFUSION BY DISCONTINUOUS MOVEMENTS, AND ON THE TELEGRAPH EQUATION},
    journal = {The Quarterly Journal of Mechanics and Applied Mathematics},
    volume = {4},
    number = {2},
    pages = {129-156},
    year = {1951},
    month = {01},
    doi = {10.1093/qjmam/4.2.129},
}

@book{leggett2006quantum,
  author    = {Leggett, Anthony James},
  title     = {Quantum Liquids: Bose Condensation and Cooper Pairing in Condensed-Matter Systems},
  publisher = {Oxford University Press},
  year      = {2006},
  series    = {Oxford Graduate Texts},
  address   = {Oxford, UK},
  isbn      = {978-0198526438}
}

@book{nozieres1999theory,
  title     = {The Theory of Quantum Liquids, Volumes I \& II},
  author    = {Nozi{\`e}res, Philippe and Pines, David},
  year      = {1999},
  publisher = {Westview Press / Avalon Publishing},
  address   = {Boulder, CO},
  isbn      = {978-0201094297}
}

@article{Huang2012,
    author = {Huang, Yonghua and Fang, Lei and Wang, Xiaojian and Wang, Ruzhu and Xu, Lie},
    title = {Thermal conductivity of helium-3 between 3 mK and 300 K},
    journal = {AIP Conference Proceedings},
    volume = {1434},
    number = {1},
    pages = {1849-1856},
    year = {2012},
    month = {06},
    issn = {0094-243X},
    doi = {10.1063/1.4707122}
}

@article{Anderson1966,
  title = {Experimental Thermal Conductivity of Helium-3},
  author = {Anderson, A. C. and Connolly, J. I. and Vilches, O. E. and Wheatley, J. C.},
  journal = {Phys. Rev.},
  volume = {147},
  issue = {1},
  pages = {86--93},
  numpages = {0},
  year = {1966},
  month = {Jul},
  publisher = {American Physical Society},
  doi = {10.1103/PhysRev.147.86},
  url = {https://aps.org}
}

@article{Mitchell1986,
  author    = {Mitchell, R. and Eastop, A. D. and Faraj, E. and Hook, J. R. and Sandiford, D. J.},
  title     = {The specific heat capacity and thermal conductivity of normal liquid {$^3$He}},
  journal   = {Journal of Low Temperature Physics},
  year      = {1986},
  volume    = {64},
  number    = {1},
  pages     = {43--63},
  doi       = {10.1007/BF00681541},
  url       = {https://doi.org/10.1007/BF00681541}
}

@article{Imry1999,
  title = {Conductance viewed as transmission},
  author = {Imry, Yoseph and Landauer, Rolf},
  journal = {Rev. Mod. Phys.},
  volume = {71},
  issue = {2},
  pages = {S306--S312},
  numpages = {0},
  year = {1999},
  month = {Mar},
  publisher = {American Physical Society},
  doi = {10.1103/RevModPhys.71.S306},
  url = {https://link.aps.org/doi/10.1103/RevModPhys.71.S306}
}

@article{Behnia2026,
      title={Deviation from {Fermi-liquid $T^2$} resistivity caused by collective transport}, 
      author={Kamran Behnia},
      year={2026},
journal = {Annalen der Physik},
volume = {538},
pages = {70301}
}

@article{BAGGIOLI20201,
title = {Gapped momentum states},
journal = {Physics Reports},
volume = {865},
pages = {1-44},
year = {2020},
note = {Gapped momentum states},
issn = {0370-1573},
doi = {https://doi.org/10.1016/j.physrep.2020.04.002},
url = {https://www.sciencedirect.com/science/article/pii/S0370157320301198},
author = {Matteo Baggioli and Mikhail Vasin and Vadim Brazhkin and Kostya Trachenko}
}

@book{ashcroft_solid_1976,
  author    = {Ashcroft, Neil W. and Mermin, N. David},
  title     = {Solid State Physics},
  publisher = {Brooks Cole},
  year      = {1976},
  edition   = {1},
  month     = {jan},
  isbn      = {0-03-083993-9},
  address   = {New York, NY}
}

@book{ziman1960electrons,
  title     = {Electrons and Phonons: The Theory of Transport Phenomena in Solids},
  author    = {Ziman, John M.},
  year      = {1960},
  publisher = {Clarendon Press},
  address   = {Oxford},
  isbn      = {9780198512356}
}

@book{abrikosov1988fundamentals,
  author    = {Abrikosov, A. A.},
  title     = {Fundamentals of the Theory of Metals},
  publisher = {North-Holland},
  address   = {Amsterdam},
  year      = {1988},
  isbn      = {978-0444870940}
}

@article{Monaco1999,
  title = {Viscoelastic behavior of water in the terahertz-frequency range: An inelastic x-ray scattering study},
  author = {Monaco, G. and Cunsolo, A. and Ruocco, G. and Sette, F.},
  journal = {Phys. Rev. E},
  volume = {60},
  issue = {5},
  pages = {5505--5521},
  numpages = {0},
  year = {1999},
  month = {Nov},
  publisher = {American Physical Society},
  doi = {10.1103/PhysRevE.60.5505},
  url = {https://link.aps.org/doi/10.1103/PhysRevE.60.5505}
}

@article{Bencivenga2007,
  title = {High-frequency dynamics of liquid and supercritical water},
  author = {Bencivenga, F. and Cunsolo, A. and Krisch, M. and Monaco, G. and Ruocco, G. and Sette, F.},
  journal = {Phys. Rev. E},
  volume = {75},
  issue = {5},
  pages = {051202},
  numpages = {11},
  year = {2007},
  month = {May},
  publisher = {American Physical Society},
  doi = {10.1103/PhysRevE.75.051202},
  url = {https://link.aps.org/doi/10.1103/PhysRevE.75.051202}
}

@article{SCOPIGNO2002121,
title = {Inelastic X-ray scattering study of the collective dynamics in simple liquid metals},
journal = {Journal of Non-Crystalline Solids},
volume = {312-314},
pages = {121-127},
year = {2002},
issn = {0022-3093},
doi = {https://doi.org/10.1016/S0022-3093(02)01660-5},
url = {https://www.sciencedirect.com/science/article/pii/S0022309302016605},
author = {T. Scopigno and U. Balucani and G. Ruocco and F. Sette}
}

@book{evans2008statistical,
  title     = {Statistical Mechanics of Nonequilibrium Liquids},
  author    = {Evans, Denis J. and Morriss, Gary P.},
  edition   = {2nd},
  year      = {2008},
  publisher = {Cambridge University Press},
  isbn      = {9781139471930},
  doi       = {10.1017/CBO9780511535307}
}

@Article{Huang2005,
author={Huang, Y. H.
and Chen, G. B.
and Li, X. Y.
and Arp, V.},
title={Density Equation for Saturated 3He},
journal={International Journal of Thermophysics},
year={2005},
month={May},
day={01},
volume={26},
number={3},
pages={729-741},
issn={1572-9567},
doi={10.1007/s10765-005-5574-6}
}

\end{document}